\documentclass[prl,twoside,twocolumn,showpacs,superscriptaddress]{revtex4}
\usepackage{graphicx}
\usepackage{amsmath}
\usepackage{amssymb}

\begin{document}

\title{Temperature and Field Dependence of the Anisotropy of MgB$_2$}
\author{M. Angst}
 \email[Email: ]{angst@phys.ethz.ch}
\affiliation{Solid State Physics Laboratory ETH, 8093 Z\"urich,
Switzerland}
\author{R. Puzniak}
\author{A. Wisniewski}
\affiliation{Institute of Physics, Polish Academy of Sciences, Al.
Lotnikow 32/46, 02-668 Warsaw, Poland}
\author{J. Jun}
\author{S. M. Kazakov}
\author{J. Karpinski}
\affiliation{Solid State Physics Laboratory ETH, 8093 Z\"urich,
Switzerland}
\author{J. Roos}
\author{H. Keller}
\affiliation{Physik-Institut, Universit\"at Z\"urich, 8057
Z\"urich, Switzerland}
\date{\today}
\begin{abstract}
The anisotropy $\gamma$ of the superconducting state of high
quality single-crystals of MgB$_2$ was determined, using torque
magnetometry with two different methods. The anisotropy of the
upper critical field was found to be temperature dependent,
decreasing from $\gamma\simeq 6$ at 15 K to 2.8 at 35 K.
Reversible torque data near $T_{\text{c}}$ reveal a field
dependent anisotropy, increasing nearly linearly from
$\gamma\simeq 2$ in zero field to $3.7$ in $10\, {\text{kOe}}$.
The unusual temperature dependence is a true bulk property and can
be explained by non-local effects of anisotropic pairing and/or
the $\vec{k}-$dependence of the effective mass tensor.
\end{abstract}
\maketitle

\label{intro} The discovery of superconductivity at
$T_{\text{c}}\approx 39 \, {\text{K}}$ in MgB$_2$
\cite{Nagamatsu01} has caused a lot of interest into it's physical
properties (for a review see Ref.\ \cite{Buzea01}). Measurements
of the isotope effect \cite{Budko01} and e.g.\ the $^{11}$B
nuclear spin-lattice relaxation rate \cite{Kotegawa01} indicated a
BCS type s-wave phonon-mediated superconductivity. Calculations of
the band structure and the phonon spectrum predict a double energy
gap \cite{Kortus01,Liu01}, a larger gap attributed to
two-dimensional $p_{x-y}$ orbitals and a smaller gap attributed to
three-dimensional $p_z$ bonding and anti-bonding orbitals. A
substantial body of evidence by, among others, specific heat
measurements \cite{Bouquet01b}, point-contact spectroscopy
\cite{Szabo01}, scanning tunneling spectroscopy \cite{Giubileo01}
and penetration depth measurements \cite{Manzano02} has emerged to
support this scenario. An alternative scenario with a single, but
anisotropic, gap was also proposed \cite{Haas01}, and supported by
recent Raman measurements \cite{Quilty02}.

A double gap structure or an anisotropic energy gap should
influence the anisotropy in the superconducting state, according
to the standard anisotropic Ginzburg-Landau (GL) theory
$\gamma=(m_c^*/m_{ab}^*)^{1/2}=\lambda_{c}/\lambda_{ab}=\xi_{ab}/\xi_{c}=H_{\text{c2}}^{\|ab}/H_{\text{c2}}^{\|c}$,
where $\|ab$($\|c$) indicates the field $H$
perpendicular(parallel) to the $c-$axis of the sample and $m^*$,
$\lambda$, $\xi$ and $H_{\text{c2}}$ are the GL effective mass,
the penetration depth, the coherence length, and the upper
critical or bulk nucleation field, respectively. Most values
reported for the anisotropy of polycrystalline or thin film
MgB$_2$ span the range of values of $\gamma=1.1-3$ \cite{Buzea01},
but there are also reports with $\gamma\approx 6-9$
\cite{Simon01,Budko01b}. Up to now, there are only four reports on
transport measurements of the upper critical field anisotropy
performed on single crystals, giving values of $2.6$ \cite{Xu01},
$2.7$ \cite{Lee01} and $3$ \cite{Kim01,Pradhan01}. Magnetic
measurements of the angular dependence of $H_{\text{c2}}(\theta)$,
yielding $\gamma=1.6$, were reported only on aligned crystallites
\cite{Lima01a}.

Here, we report magnetic torque measurements on single crystals of
MgB$_2$, performed in a wide range of temperatures from $15\,
{\text{K}}$ to $36\, {\text{K}}$ in magnetic fields of up to $90\,
{\text{kOe}}$. We provide evidence that the bulk anisotropy
$\gamma$ is not universally constant, but is temperature dependent
down to at least $0.4\, T_{\text{c}}$ and shows a pronounced field
dependence near $T_{\text{c}}$. Microscopic origins of the unusual
$T-$dependence of $\gamma$ in MgB$_2$ are discussed.

\label{exp} We have grown single crystals of MgB$_2$ with a high
pressure cubic anvil technique, similar to the one described in
Ref.\ \cite{Lee01}. The details of crystal growth will be
published elsewhere. In brief, a mixture of Mg and B was put into
a BN container and a pressure of $30-35\, {\text{kbar}}$ was
applied. Growth runs consisted of heating during $1 \, {\text{h}}$
up to the maximum temperature of $1700-1800^{\circ}{\text{C}}$,
keeping the temperature for $1-3\, {\text{h}}$ and then cooling to
room temperature during $1-2\, {\text{h}}$. Flat crystals were up
to $0.8\times 0.6\times 0.05 \, {\text{mm}}^3$ in size, with sharp
transitions to the superconducting state at about $38-39\,
{\text{K}}$.

Measurements were performed on miniaturized piezoresistive
cantilevers specifically designed for torque magnetometry
\cite{Willemin98b}. The torque $\vec{\tau}=\vec{m}\times
\vec{B}\simeq \vec{m}\times \vec{H}$, where $\vec{m}$ is the
magnetic moment of the sample, was recorded as a function of the
angle $\theta$ between the applied field $\vec{H}$ and the
$c-$axis of the crystal for various fixed temperatures and fields.
For measurements close to $T_{\text{c}}$, in fields up to $14 \,
{\text{kOe}}$, a non-commercial magnetometer with very high
sensitivity was used. For part of these measurements, a
vortex-shaking process was employed to speed up the relaxation of
the vortex lattice \cite{Willemin98}. The observation of a
well-resolved lock-in effect in $\tau(\theta)$ (see upper inset of
Fig.\ \ref{Fig5}) indicates there are no variations of
crystallographic alignment throughout the samples. A crystal with
a volume of about $4\times 10^{-4}\, {\text{mm}}^3$ (sample A) was
measured in this system. Another crystal with a volume of about
$8\times 10^{-3}\, {\text{mm}}^3$ (sample B) was measured in a
wider range of temperatures down to $15\, {\text{K}}$ in a Quantum
Design PPMS with torque option and a maximum field of $90\,
{\text{kOe}}$.

\label{res}
\begin{figure}[tb]
\includegraphics[width=0.95\linewidth]{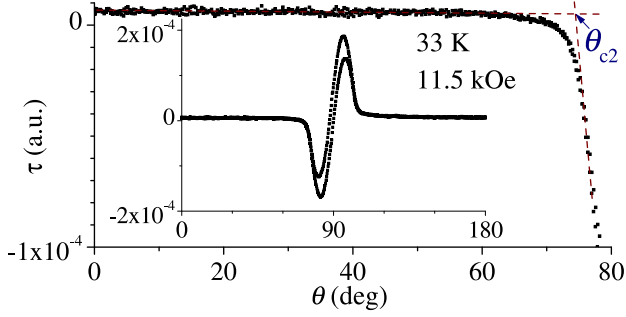}
\caption{Torque $\tau$ vs.\ angle $\theta$ data, showing the
definition of $\theta_{\text{c2}}$. The inset shows the full
torque curve measured. } \label{Fig1}
\end{figure}

An example of a torque vs.\ angle curve is given in the inset of
Fig.\ \ref{Fig1}. For small angles $\theta$ the torque is
essentially zero. Only when $H$ is nearly parallel to the
$ab-$plane there is an appreciable torque signal. The curve can be
interpreted in a straight-forward way: for $H$ parallel to the
$c-$axis the sample is in the normal state, while for $H$ parallel
to the $ab-$plane it is in the superconducting state. The
crossover angle $\theta_{\text{c2}}$ between the normal and the
superconducting state is the angle for which the fixed applied
field is the upper critical field. The inset of Fig.\ \ref{Fig1}
also shows hysteretic behaviour due to irreversibility. The
irreversibility field $H_{\text{irr}}(T,\theta)$ determined from
the torque measurements is very high, close to $H_{\text{c2}}$.
Preliminary SQUID measurements on similar crystals indicate a much
lower $H_{\text{irr}}$ \cite{HirrTS}; an extended discussion of
the irreversible properties of MgB$_2$ will be published
elsewhere.

\begin{figure}[tb]
\includegraphics[width=0.95\linewidth]{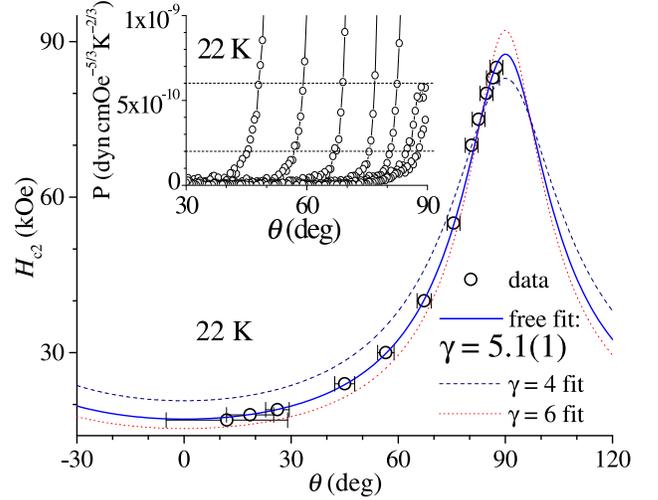}
\caption{Upper critical field $H_{\text{c2}}$ vs.\ $\theta$ at
$22\ {\text{K}}$. The full line is a free fit of Eq.\
(\ref{Hc2_theta}) to the data. Alternative fits with fixed values
of the anisotropy $\gamma$ are also shown. The inset shows
selected rescaled torque P (see Eq.\ (\ref{scaling})) vs.\ angle
$\theta$. From left to right, curves shown were measured in
$H=24,30,40,55,70,80,$ and $85\, \text{kOe}$. The two criteria
used for the determination of $\theta_{\text{c2}}$ are indicated
by dashed lines. The data in the main panel were obtained
employing the lower criterion.} \label{Fig2}
\end{figure}

The crossing between straight lines through the background and the
superconducting torque signal was used to define
$\theta_{\text{c2}}$. This definition is not unambiguous. Taking
the analysis of the data more strict we have to apply the
appropriate scaling rules. The magnetization $M$ of a 3D system in
the GL theory of fluctuations in the vicinity of the transition
temperature $T_{\text{c}}(H)$ in high magnetic fields is given by
a universal function F of the distance from $T_{\text{c}}(H)$
\cite{Welp91}:
\begin{equation}
\frac{M}{H} = \frac{T^{2/3}}{H^{1/3}}\text{F}\left(
\frac{{\text{A}}(T-T_{\text{c}})}{(TH)^{2/3}}\right),
\label{MHscal}
\end{equation}
where A is a material constant. Combining the above dependence
with the angular dependence of the torque \cite{Buzdin92} we find
that the rescaled torque signal
\begin{equation}
\text{P}=-\tau\epsilon^{1/3}(\theta) \left/ \left(
\sin\theta\cos\theta H^{5/3}(1-1/\gamma^2)T^{2/3} \right) \right.
 \label{scaling}
\end{equation}
with
$\epsilon(\theta)=(\cos^2\theta+\sin^2\theta/\gamma^2)^{1/2}$, is
a universal function of the distance from $T_{\text{c}}$ with a
fixed value $\text{F}(0)$ at $T=T_{\text{c}}(H)$. Taking into
account the $\text{F}(0)$ value for the theoretical dependence of
the universal function for a 3D system \cite{Wilkin93} we can
estimate that for a volume of the sample of $8\times 10^{-3}\,
{\text{mm}}^3$  P reaches at $T=T_{\text{c}}(H)$ a value of about
$2\times 10^{-10}\, {\text{dyn}}\,
{\text{cmOe}}^{-5/3}{\text{K}}^{-2/3}$. The inset in Fig.\
\ref{Fig2} presents the angular dependence of the rescaled torque
P in different magnetic fields at 22K. The crossing of the
${\text{P}}(\theta)$ dependence for each field with the line of
the constant value of $2\times 10^{-10}\, {\text{dyn}}\,
{\text{cmOe}}^{-5/3}{\text{K}}^{-2/3}$ determines the
$H_{\text{c2}}(\theta)$ dependence as it is shown in the main
panel of Fig.\ \ref{Fig2}. It is important to stress that the
results obtained depend not very sensitively on the criterion
chosen and it will be shown later (see Fig.\ \ref{Fig4}) that even
with a three times higher criterion we get very similar
temperature dependences of $H_{\text{c2}}$ and $\gamma$.
Additional $\tau(H)$ measurements at fixed angle give
$H_{\text{c2}}(\theta)$ values corresponding well to those from
$\tau(\theta)$ measurements.

\begin{figure}[t!b]
\includegraphics[width=0.95\linewidth]{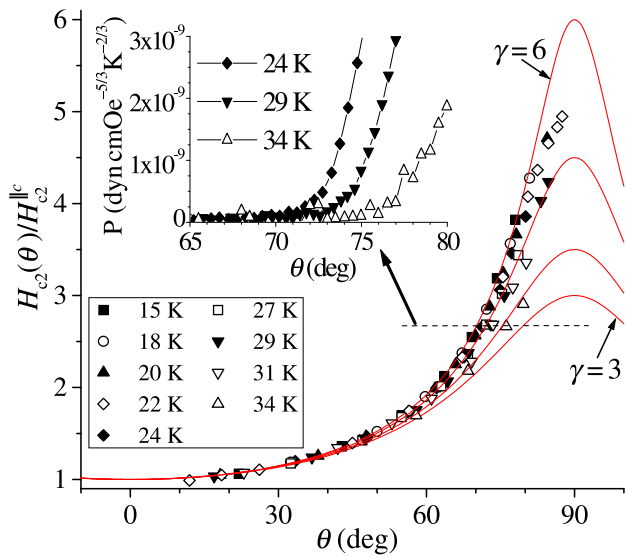}
\caption{Angular dependence of
$H_{\text{c2}}(\theta)/H_{\text{c2}}^{\|c}$, employing the lower
criterion, at various temperatures. Curves correspond to
$H_{\text{c2}}(\theta)$ according to Eq.\ (\ref{Hc2_theta}) for
$\gamma=3,3.5,4.5,$ and $6$. Inset: Some rescaled torque P (see
Eq.\ (\ref{scaling})) vs.\ angle $\theta$ curves, corresponding to
$H/H_{\text{c2}}^{\|c}\simeq 2.67$.} \label{Fig3}
\end{figure}

Within the applicability of the anisotropic GL theory the angle
dependence of the upper critical field is predicted to be
\cite{Tilley65}
\begin{equation}
H_{\text{c2}}(\theta)=H_{\text{c2}}^{\|c} \left( \cos^2 \theta +
\sin^2 \theta /\gamma^2 \right)^{-1/2}. \label{Hc2_theta}
\end{equation}
A fit of Eq.\ (\ref{Hc2_theta}) to the data at $22 \, {\text{K}}$
yields $\gamma=5.1(1)$ and $H_{\text{c2}}^{\|c} = 17.2(1) \,
{\text{kOe}}$. This fit ($\gamma=5.1$) describes the data well,
while alternative fits with $\gamma$ fixed to $4$ and $6$ are
clearly incompatible with the data, as shown in Fig.\ \ref{Fig2}.

Figure \ref{Fig3} shows the angular dependence of $H_{\text{c2}}$
scaled by $H_{\text{c2}}^{\|c}$ to directly compare the anisotropy
at different temperatures. The $15\, {\text{K}}$ data are well
described by the line corresponding to $\gamma=6$, while the $34\,
{\text{K}} $ data lie below the line for $\gamma=3.5$. The data
indicate an anisotropy systematically decreasing with increasing
temperature. To show this is not an artifact related to fitting,
we present the angular dependence of the rescaled torque P for
fixed $H/H_{\text{c2}}^{\|c}$ in the inset. The curves clearly
shift to higher angles with increasing temperature. Furthermore,
we directly checked $\tau(\theta)$ raw data in fields above and
below $H_{\text{c2}}^{\|c}$ and $H_{\text{c2}}^{\|ab}$ to give
absolute limitations of $\gamma$. We find that at $22\,
{\text{K}}$, the anisotropy {\em{must}} be higher than $4.4$,
while at $34\, {\text{K}}$, it {\em{must}} be lower than $3.5$.

\begin{figure}[b!]
\includegraphics[width=0.95\linewidth]{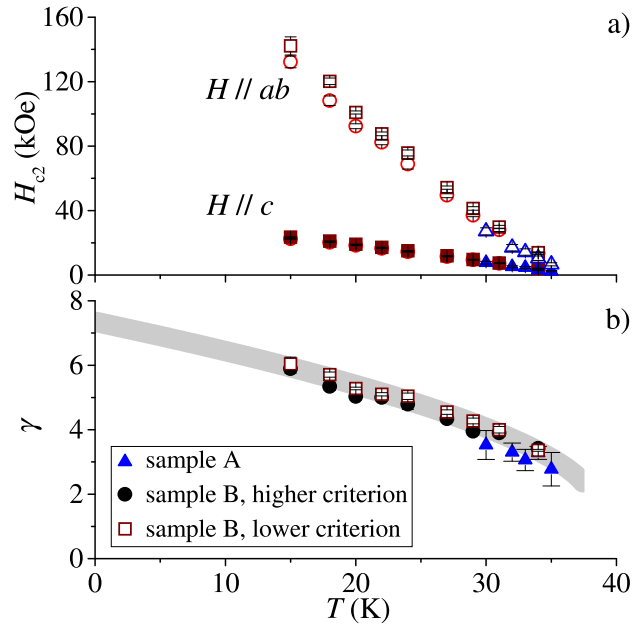}
\caption{a) Upper critical field $H_{\text{c2}}$ vs.\ temperature
$T$. Open symbols correspond to $H\parallel ab$, full symbols to
$H\parallel c$, from fits of $H_{\text{c2}}(\theta)$ data to Eq.\
(\ref{Hc2_theta}). Up triangles are from measurements on sample A
(with $\theta_{\text{c2}}$ determined as shown in Fig.\
\ref{Fig1}) and squares (circles) are from measurements on sample
B, using the lower (higher) criterion of constant rescaled torque
P (see inset of Fig.\ \ref{Fig2}). b) Temperature dependence of
the upper critical field anisotropy
$H_{\text{c2}}^{\|ab}/H_{\text{c2}}^{\|c}$, determined from fits
of $H_{\text{c2}}(\theta)$ to Eq.\ (\ref{Hc2_theta}).}
\label{Fig4}
\end{figure}
\begin{figure}[tb]
\includegraphics[width=0.95\linewidth]{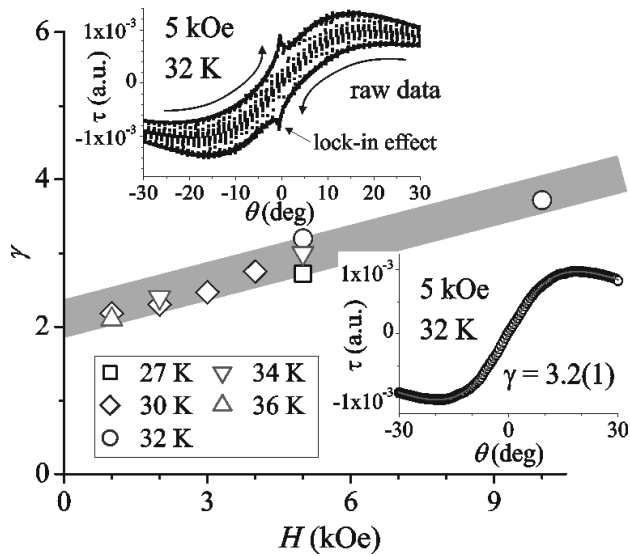}
\caption{Field dependence of the anisotropy $\gamma$ as determined
from the angular dependence of the reversible torque
$\tau(\theta)$. Upper inset: raw data, demonstrating the effect of
shaking and the lock-in effect. Lower inset: fitting of the shaked
data to a formula derived by Kogan {\em{et al.}} \cite{Kogan88},
yielding $\gamma\simeq 3.2$.} \label{Fig5}
\end{figure}

All data are summarized in Fig.\ \ref{Fig4}. The
$H_{\text{c2}}^{\|c}$ data obtained from fits to Eq.\
(\ref{Hc2_theta}) do not vary much with the criterion used for the
determination of $\theta_{\text{c2}}$, and agree well with
$H_{\text{c2}}^{\|c}$ calculated from thermal conductivity data
\cite{Sologubenko_pc} measured on a single crystal grown with the
same technique. The $T-$dependence of $H_{\text{c2}}^{\|c}$ is in
agreement with calculations by Helfand {\em{et al.}}
\cite{Helfand66}. The corresponding $H_{\text{c2}}^{\|c}(0)\simeq
31\, {\text{kOe}}$ is relatively small compared to literature
values, which may indicate that the crystals investigated are
relatively free of defects. The $H_{\text{c2}}^{\|ab}$ values were
obtained from the two fit parameters $H_{\text{c2}}^{\|c}$ and
$\gamma$. There is a slight positive curvature of
$H_{\text{c2}}^{\|ab}(T)$, which can be attributed to the
$T-$dependence of $\gamma$. The $\gamma(T)$ dependence may also be
the origin of the positive curvature of $H_{\text{c2}}$ observed
in other measurements of bulk, thin film and single crystal
measurements \cite{Buzea01}. Due to the lack of low temperature
data and the variation of $\gamma$, only a rough estimation
$H_{\text{c2}}^{\|ab}(0)\approx 230\, {\text{kOe}}$ can be given.
The anisotropy data show that $\gamma$ {\em{systematically}}
{\em{decreases}} with increasing temperature. A change of the
criterion used for the determination of $\theta_{\text{c2}}$ leads
to small shifts of the magnitude of $\gamma$, but the temperature
dependence is always the same. The highest upper critical field
anisotropy $\gamma\simeq 6$ was obtained at $15\, {\text{K}}$, the
lowest anisotropy $\gamma\simeq 2.8$ at $35\, {\text{K}}$. From
Fig.\ \ref{Fig4} we estimate $\gamma(0) = 7-8$ and
$\gamma(T_{\text{c}}) = 2.3-2.7$.

Small systematic deviations from Eq.\ (\ref{Hc2_theta}), observed
near $T_{\text{c}}$, indicate that the field influence on $\gamma$
may be important as well. To clarify this point, we have measured
the reversible torque $\tau$ as a function of angle $\theta$ for
various fields and temperatures near $T_{\text{c}}$. The data were
analyzed with an equation derived by Kogan {\em{et al.}}
\cite{Kogan88}, based on the anisotropic London model, which
contains the GL anisotropy $\gamma$ as a parameter.

Without shaking, the irreversibility of the torque was relatively
large and a clear lock-in effect was observed. The inset of Fig.\
\ref{Fig5} shows an example of shaked torque data, together with
the fitted curve. An evaluation of the data measured for various
$T$ and $H$ up to $10\, {\text{kOe}}$ with the Kogan formula
\cite{Kogan88} reveal that $\gamma$ is field-dependent with larger
$\gamma$ in larger fields, while temperature variations do not
affect $\gamma$ appreciably (see Fig.\ \ref{Fig5}). Indications of
a field dependence of the coherence length $\xi$ (a precondition
of a field dependence of it's anisotropy) have been observed by
specific heat \cite{Sonier99} and muon spin rotation
\cite{Sonier97} measurements on NbSe$_2$. We should stress that
Kogan's formula assumes equivalence of the anisotropies of the
coherence length $\xi$ and the penetration depth $\lambda$. Since
the anisotropy of $\lambda$ might be different from the one of
$\xi$ in our case, the two different methods used in this work can
lead to different effective values of $\gamma$. The field
dependence of $\gamma$ may be related to the peculiar double gap
structure of MgB$_2$ with a large gap of two-dimensional nature
and a small three-dimensional gap, which is very rapidly
suppressed in a magnetic field \cite{Szabo01}.

A temperature dependent $H_{\text{c2}}$ anisotropy was previously
observed, e.g.\ in NbSe$_2$ \cite{Muto77} and LuNi$_2$B$_2$C
\cite{Metlushko97}. However, in MgB$_2$, the effect is much more
pronounced. It was shown that any theory capable of explaining a
temperature dependence of $\gamma$, needs to take into account
non-local effects \cite{Takanaka75,Teichler75}, which can be
pronounced in samples of high purity. In the vicinity of
$T_{\text{c}}$, non-locality is not important \cite{Metlushko97}.
Therefore, $m_c^*/m_{ab}^* = \gamma^2(T_{\text{c}})\simeq 5-7$
corresponds to the standard GL effective mass anisotropy. We note
that this is significantly higher than the calculated
\cite{Kortus01,Belashchenko01} anisotropy of the band effective
mass (BEM) averaged over the Fermi surface ($1.0-1.2$). The
microscopic theories \cite{Takanaka75,Teichler75} show that the
$T-$dependence of $\gamma$ cannot be attributed to an anisotropy
of the BEM tensor, unless it is also wave vector dependent. An
anisotropic energy gap, caused e.g.\ by an anisotropy in the
electron-phonon coupling (EPC), can also lead to variations of
$\gamma$ with temperature.

In MgB$_2$, first principles calculations suggest both a
pronounced wave-vector dependence of the BEM tensor
\cite{Belashchenko01} and a strong anisotropy of the EPC (see
e.g.\ \cite{An01,Liu01}). The latter leads naturally \cite{Liu01}
to the observed double energy gap and was suggested \cite{Choi01}
to be responsible for the unusually high $T_{\text{c}}$ of
MgB$_2$. To our knowledge, there is only one theory calculating a
temperature dependent $H_{\text{c2}}$ anisotropy of MgB$_2$
\cite{Haas01}, which predicts, however,
$H_{\text{c2}}^{\|ab}/H_{\text{c2}}^{\|c}$ to increase with
increasing $T$. A quantitative explanation of the measured
$\gamma(T)$ apparently needs to take into account both the
wave-vector dependence of the BEM tensor and the anisotropic EPC,
and is beyond the scope of this work.

\label{conc} In conclusion, the upper critical field anisotropy
$\gamma$ of MgB$_2$, determined by torque magnetometry, decreases
with increasing temperature. Measurements of the reversible torque
near $T_{\text{c}}$ reveal an almost linear field dependence of
the anisotropy of the coherence length and/or the penetration
depth as well. Our results imply a breakdown of standard
anisotropic GL theory with a (temperature and field independent)
effective mass anisotropy. The temperature dependence of $\gamma$
can be approximated tentatively by
$\gamma(T)=\gamma^{*}+\tilde{\gamma} (1-T/T_{\text{c}})^n$ with
$n$ close to $1$. Here, $\gamma^{*}\simeq 2.3-2.7$ is the band
effective mass anisotropy and $\tilde{\gamma} \simeq 4.5-5.5$
arises from the anisotropy of the attractive electron-electron
interaction and/or the wave-vector dependence of the effective
mass tensor.

\begin{acknowledgments}
\label{ack} We thank B.~Batlogg, P.~{Miranovi\'{c}},
A.~Sologubenko, and I.~L. Landau for useful discussions. This work
was supported by the Swiss National Science Foundation, by the
European Community (contract ICA1-CT-2000-70018) and by the Polish
State Committee for Scientific Research (5 P03B 12421).
\end{acknowledgments}

\newcommand{\noopsort}[1]{} \newcommand{\printfirst}[2]{#1}
  \newcommand{\singleletter}[1]{#1} \newcommand{\switchargs}[2]{#2#1}


\begin{thebibliography}{36}
\expandafter\ifx\csname
natexlab\endcsname\relax\def\natexlab#1{#1}\fi
\expandafter\ifx\csname bibnamefont\endcsname\relax
  \def\bibnamefont#1{#1}\fi
\expandafter\ifx\csname bibfnamefont\endcsname\relax
  \def\bibfnamefont#1{#1}\fi
\expandafter\ifx\csname citenamefont\endcsname\relax
  \def\citenamefont#1{#1}\fi
\expandafter\ifx\csname url\endcsname\relax
  \def\url#1{\texttt{#1}}\fi
\expandafter\ifx\csname
urlprefix\endcsname\relax\def\urlprefix{URL }\fi
\providecommand{\bibinfo}[2]{#2}
\providecommand{\eprint}[2][]{\url{#2}}

\bibitem[{\citenamefont{Nagamatsu et~al.}(2001)\citenamefont{Nagamatsu,
  Nakagawa, Muranaka, Zenitani, and Akimitsu}}]{Nagamatsu01}
\bibinfo{author}{\bibfnamefont{J.}~\bibnamefont{Nagamatsu}}
  \bibnamefont{{\em{et~al.}},}
  \bibinfo{journal}{Nature} \textbf{\bibinfo{volume}{410}}, \bibinfo{pages}{63}
  (\bibinfo{year}{2001}).

\bibitem[{\citenamefont{Buzea and Yamashita}(2001)}]{Buzea01}
\bibinfo{author}{\bibfnamefont{C.}~\bibnamefont{Buzea}} \bibnamefont{and}
  \bibinfo{author}{\bibfnamefont{T.}~\bibnamefont{Yamashita}},
  \bibinfo{journal}{Supercond. Sci. Technol.} \textbf{\bibinfo{volume}{14}},
  \bibinfo{pages}{R115} (\bibinfo{year}{2001}).

\bibitem[{\citenamefont{Bud'ko et~al.}(2001{\natexlab{a}})\citenamefont{Bud'ko,
  Lapertot, Petrovic, Cunningham, Anderson, and Canfield}}]{Budko01}
\bibinfo{author}{\bibfnamefont{S.~L.} \bibnamefont{Bud'ko}}
  \bibnamefont{{\em{et~al.}},}
  \bibinfo{journal}{Phys. Rev. Lett.} \textbf{\bibinfo{volume}{86}},
  \bibinfo{pages}{1877} (\bibinfo{year}{2001}{\natexlab{a}}).

\bibitem[{\citenamefont{Kotegawa et~al.}(2001)\citenamefont{Kotegawa, Ishida,
  Kitaoka, Muranaka, and Akimitsu}}]{Kotegawa01}
\bibinfo{author}{\bibfnamefont{H.}~\bibnamefont{Kotegawa}}
   \bibnamefont{{\em{et~al.}},}
  \bibinfo{journal}{Phys. Rev. Lett.} \textbf{\bibinfo{volume}{87}},
  \bibinfo{pages}{127001} (\bibinfo{year}{2001}).

\bibitem[{\citenamefont{Kortus et~al.}(2001)\citenamefont{Kortus, Mazin,
  Belashchenko, Antropov, and Boyer}}]{Kortus01}
\bibinfo{author}{\bibfnamefont{J.}~\bibnamefont{Kortus}}
  \bibnamefont{{\em{et~al.}},}
  \bibinfo{journal}{Phys. Rev. Lett.} \textbf{\bibinfo{volume}{86}},
  \bibinfo{pages}{4656} (\bibinfo{year}{2001}).

\bibitem[{\citenamefont{Liu et~al.}(2001)\citenamefont{Liu, Mazin, and
  Kortus}}]{Liu01}
\bibinfo{author}{\bibfnamefont{A.~Y.} \bibnamefont{Liu}},
  \bibinfo{author}{\bibfnamefont{I.~I.} \bibnamefont{Mazin}}, \bibnamefont{and}
  \bibinfo{author}{\bibfnamefont{J.}~\bibnamefont{Kortus}},
  \bibinfo{journal}{Phys. Rev. Lett.} \textbf{\bibinfo{volume}{87}},
  \bibinfo{pages}{087005} (\bibinfo{year}{2001}).

\bibitem[{\citenamefont{Bouquet et~al.}(2001)\citenamefont{Bouquet, Fisher,
  Phillips, Hinks, and Jorgensen}}]{Bouquet01b}
\bibinfo{author}{\bibfnamefont{F.}~\bibnamefont{Bouquet}}
  \bibnamefont{{\em{et~al.}},}
  \bibinfo{journal}{Phys. Rev. Lett.} \textbf{\bibinfo{volume}{87}},
  \bibinfo{pages}{047001} (\bibinfo{year}{2001}).

\bibitem[{\citenamefont{Szab{\'o} et~al.}(2001)\citenamefont{Szab{\'o},
  Samuely, Kamar{\'{\i}}k, Klein, Marcus, Fruchart, Miraglia, Marcenat, and
  Jansen}}]{Szabo01}
\bibinfo{author}{\bibfnamefont{P.}~\bibnamefont{Szab{\'o}}}
  \bibnamefont{{\em{et~al.}},}
  \bibinfo{journal}{Phys. Rev. Lett.} \textbf{\bibinfo{volume}{87}},
  \bibinfo{pages}{137005} (\bibinfo{year}{2001}).

\bibitem[{\citenamefont{Giubileo et~al.}(2001)\citenamefont{Giubileo,
  Roditchev, Sacks, Lamy, and Klein}}]{Giubileo01}
\bibinfo{author}{\bibfnamefont{F.}~\bibnamefont{Giubileo}}
  \bibnamefont{{\em{et~al.}},}
  \bibinfo{journal}{Phys. Rev. Lett.} \textbf{\bibinfo{volume}{87}},
  \bibinfo{pages}{177008} (\bibinfo{year}{2001}).

\bibitem[{\citenamefont{Manzano et~al.}(2002)\citenamefont{Manzano,
  Carrington, Hussey, Lee, Yamamoto, and Tajima}}]{Manzano02}
\bibinfo{author}{\bibfnamefont{F.}~\bibnamefont{Manzano}}
  \bibnamefont{{\em{et~al.}},}
  \bibinfo{journal}{Phys. Rev. Lett.} \textbf{\bibinfo{volume}{88}},
  \bibinfo{pages}{047002} (\bibinfo{year}{2002}).

\bibitem[{\citenamefont{Haas and Maki}(2001)}]{Haas01}
\bibinfo{author}{\bibfnamefont{S.}~\bibnamefont{Haas}} \bibnamefont{and}
  \bibinfo{author}{\bibfnamefont{K.}~\bibnamefont{Maki}},
  \bibinfo{journal}{Phys. Rev. B} \textbf{\bibinfo{volume}{65}},
  \bibinfo{pages}{020502(R)} (\bibinfo{year}{2002}).

\bibitem[{\citenamefont{Quilty et~al.}(2002)\citenamefont{Quilty,
  Lee, Yamamoto, and Tajima}}]{Quilty02}
\bibinfo{author}{\bibfnamefont{J.~W.} \bibnamefont{Quilty}}
  \bibnamefont{{\em{et~al.}},}
  \bibinfo{journal}{Phys. Rev. Lett.} \textbf{\bibinfo{volume}{88}},
  \bibinfo{pages}{087001} (\bibinfo{year}{2002}).

\bibitem[{\citenamefont{Simon et~al.}(2001)\citenamefont{Simon, J{\'a}nossy,
  Feh{\'e}r, Mur{\'a}nyi, Garaj, Forr{\'o}, Petrovic, L.{ Bud'ko}, Lapertot,
  Kogan et~al.}}]{Simon01}
\bibinfo{author}{\bibfnamefont{F.}~\bibnamefont{Simon}}
  \bibnamefont{{\em{et~al.}},}
  \bibinfo{journal}{Phys. Rev. Lett.}
  \textbf{\bibinfo{volume}{87}}, \bibinfo{pages}{047002}
  (\bibinfo{year}{2001}).

\bibitem[{\citenamefont{Bud'ko et~al.}(2001{\natexlab{b}})\citenamefont{Bud'ko,
  Kogan, and Canfield}}]{Budko01b}
\bibinfo{author}{\bibfnamefont{S.~L.} \bibnamefont{Bud'ko}},
  \bibinfo{author}{\bibfnamefont{V.~G.} \bibnamefont{Kogan}}, \bibnamefont{and}
  \bibinfo{author}{\bibfnamefont{P.~C.} \bibnamefont{Canfield}},
  \bibinfo{journal}{Phys. Rev. B}
  \textbf{\bibinfo{volume}{64}}, \bibinfo{pages}{180506}
  (\bibinfo{year}{2001}).

\bibitem[{\citenamefont{Xu et~al.}(2001)\citenamefont{Xu, Kitazawa, Takano, Ye,
  Nishida, Abe, Matsushita, and Kido}}]{Xu01}
\bibinfo{author}{\bibfnamefont{M.}~\bibnamefont{Xu}}
  \bibnamefont{{\em{et~al.}},}
  \bibinfo{journal}{Appl. Phys. Lett.} \textbf{\bibinfo{volume}{79}},
  \bibinfo{pages}{2779} (\bibinfo{year}{2001}).

\bibitem[{\citenamefont{S.Lee et~al.}(2001)\citenamefont{S.Lee, H.Mori,
  T.Masui, Eltsev, A.Yamamoto, and S.Tajima}}]{Lee01}
\bibinfo{author}{\bibnamefont{S. Lee}}
  \bibnamefont{{\em{et~al.}},}
  \bibinfo{journal}{J. Phys. Soc.
  Jpn.} \textbf{\bibinfo{volume}{70}}, \bibinfo{pages}{2255}
  (\bibinfo{year}{2001}).

\bibitem[{\citenamefont{Kim et~al.}(2001)\citenamefont{Kim, Choi, Jung,
  Chowdhury, Park, Kim, Kim, Du, Choi, Kim et~al.}}]{Kim01}
\bibinfo{author}{\bibfnamefont{K.~H.~P.} \bibnamefont{Kim}}
  \bibnamefont{{\em{et~al.}},}
  \bibinfo{journal}{Phys. Rev. B}
  \textbf{\bibinfo{volume}{65}}, \bibinfo{pages}{100510(R)}
  (\bibinfo{year}{2001}).

\bibitem[{\citenamefont{Pradhan et~al.}(2001)\citenamefont{Pradhan, Shi,
  Tokunaga, Takano, Togano, Kito, and Ihara}}]{Pradhan01}
\bibinfo{author}{\bibfnamefont{A.~K.} \bibnamefont{Pradhan}}
   \bibnamefont{{\em{et~al.}},}
  \bibinfo{journal}{Phys. Rev. B} \textbf{\bibinfo{volume}{64}},
  \bibinfo{pages}{212509} (\bibinfo{year}{2001}).

\bibitem[{\citenamefont{de~Lima et~al.}(2001)\citenamefont{de~Lima, Cardoso,
  Ribeiro, Avila, and Coelho}}]{Lima01a}
\bibinfo{author}{\bibfnamefont{O.~F.} \bibnamefont{de~Lima}}
  \bibnamefont{{\em{et~al.}},}
  \bibinfo{journal}{Phys. Rev. B} \textbf{\bibinfo{volume}{64}},
  \bibinfo{pages}{144517} (\bibinfo{year}{2001}).

\bibitem[{\citenamefont{Willemin
  et~al.}(1998{\natexlab{a}})\citenamefont{Willemin, Rossel, Brugger, Despont,
  Rothuizen, Vettiger, Hofer, and Keller}}]{Willemin98b}
\bibinfo{author}{\bibfnamefont{M.}~\bibnamefont{Willemin}}
  \bibnamefont{{\em{et~al.}},}
  \bibinfo{journal}{J. Appl. Phys.} \textbf{\bibinfo{volume}{83}},
  \bibinfo{pages}{1163} (\bibinfo{year}{1998}{\natexlab{a}}).

\bibitem[{\citenamefont{Willemin
  et~al.}(1998{\natexlab{b}})\citenamefont{Willemin, Rossel, Hofer, Keller,
  Erb, and Walker}}]{Willemin98}
\bibinfo{author}{\bibfnamefont{M.}~\bibnamefont{Willemin}}
  \bibnamefont{{\em{et~al.}},}
  \bibinfo{journal}{Phys. Rev. B} \textbf{\bibinfo{volume}{58}},
  \bibinfo{pages}{R5940} (\bibinfo{year}{1998}{\natexlab{b}}).

\bibitem{HirrTS} Higher irreversibility fields from torque
measurements than from SQUID measurements on the same samples were
often observed before. See, e.g., D.~Zech {\em{et~al.}}, Phys.
Rev. B \textbf{54}, 6129 (1996).

\bibitem[{\citenamefont{Welp et~al.}(2001)\citenamefont{Welp, Fleshler,
  Kwok, Klemm, Vinokur, Downey, Veal, and Crabtree}}]{Welp91}
\bibinfo{author}{\bibfnamefont{U.}~\bibnamefont{Welp}}
  \bibnamefont{{\em{et~al.}},}
  \bibinfo{journal}{Phys. Rev. Lett.} \textbf{\bibinfo{volume}{67}},
  \bibinfo{pages}{3180} (\bibinfo{year}{1991}).

\bibitem[{\citenamefont{Buzdin and Feinberg}(1992)}]{Buzdin92}
\bibinfo{author}{\bibfnamefont{A.}~\bibnamefont{Buzdin}} \bibnamefont{and}
  \bibinfo{author}{\bibfnamefont{D.}~\bibnamefont{Feinberg}},
  \bibinfo{journal}{Physica C} \textbf{\bibinfo{volume}{220}},
  \bibinfo{pages}{74} (\bibinfo{year}{1992}).

\bibitem[{\citenamefont{Wilkin and Moore}(1993)}]{Wilkin93}
\bibinfo{author}{\bibfnamefont{N.~K.} \bibnamefont{Wilkin}} \bibnamefont{and}
  \bibinfo{author}{\bibfnamefont{M.~A.} \bibnamefont{Moore}},
  \bibinfo{journal}{Phys. Rev. B} \textbf{\bibinfo{volume}{48}},
  \bibinfo{pages}{3464} (\bibinfo{year}{1993}).

\bibitem[{\citenamefont{Tilley}(1965)}]{Tilley65}
\bibinfo{author}{\bibfnamefont{D.~R.} \bibnamefont{Tilley}},
  \bibinfo{journal}{Proc. Phys. Soc. London} \textbf{\bibinfo{volume}{86}},
  \bibinfo{pages}{289} (\bibinfo{year}{1965}).

\bibitem[{\citenamefont{Kogan et~al.}(1988)\citenamefont{Kogan, Fang, and
  Mitra}}]{Kogan88}
\bibinfo{author}{\bibfnamefont{V.~G.} \bibnamefont{Kogan}},
  \bibinfo{author}{\bibfnamefont{M.~M.} \bibnamefont{Fang}}, \bibnamefont{and}
  \bibinfo{author}{\bibfnamefont{S.}~\bibnamefont{Mitra}},
  \bibinfo{journal}{Phys. Rev. B} \textbf{\bibinfo{volume}{38}},
  \bibinfo{pages}{R11958} (\bibinfo{year}{1988}).

\bibitem[{\citenamefont{Sologubenko, Jun, Kazakov, Karpinski, and Ott}}]{Sologubenko_pc}
\bibinfo{author}{\bibfnamefont{A.}~\bibnamefont{Sologubenko}}
  \bibnamefont{{\em{et~al.}},}
  (\bibinfo{year}{2001}), cond-mat/0112191.

\bibitem[{\citenamefont{Helfand and Werthamer}(1966)}]{Helfand66}
\bibinfo{author}{\bibfnamefont{E.}~\bibnamefont{Helfand}} \bibnamefont{and}
  \bibinfo{author}{\bibfnamefont{N.~R.} \bibnamefont{Werthamer}},
  \bibinfo{journal}{Phys. Rev.} \textbf{\bibinfo{volume}{147}},
  \bibinfo{pages}{288} (\bibinfo{year}{1966}).

\bibitem[{\citenamefont{Sonier et~al.}(1999)\citenamefont{Sonier, Hundley,
  Thompson, and Brill}}]{Sonier99}
\bibinfo{author}{\bibfnamefont{J.~E.} \bibnamefont{Sonier}}
   \bibnamefont{{\em{et~al.}},}
  \bibinfo{journal}{Phys. Rev. Lett.} \textbf{\bibinfo{volume}{82}},
  \bibinfo{pages}{4914} (\bibinfo{year}{1999}).

\bibitem[{\citenamefont{Sonier et~al.}(1997)\citenamefont{Sonier, Kiefl,
  Brewer, Chakhalian, Dunsiger, MacFarlane, Miller, Wong, Luke, and
  Brill}}]{Sonier97}
\bibinfo{author}{\bibfnamefont{J.~E.} \bibnamefont{Sonier}}
  \bibnamefont{{\em{et~al.}},}
  \bibinfo{journal}{Phys. Rev. Lett.} \textbf{\bibinfo{volume}{79}},
  \bibinfo{pages}{1742} (\bibinfo{year}{1997}).

\bibitem[{\citenamefont{Muto et~al.}(1977)\citenamefont{Muto, Noto, Nakatsuji,
  and Toyota}}]{Muto77}
\bibinfo{author}{\bibfnamefont{Y.}~\bibnamefont{Muto}}
  \bibnamefont{{\em{et~al.}},}
  \bibinfo{journal}{Nuovo Cimento B} \textbf{\bibinfo{volume}{38}},
  \bibinfo{pages}{503} (\bibinfo{year}{1977}).

\bibitem[{\citenamefont{Metlushko et~al.}(1997)\citenamefont{Metlushko,
  Welp, Koshelev, Aranson, and Crabtree}}]{Metlushko97}
\bibinfo{author}{\bibfnamefont{V.}~\bibnamefont{Metlushko}}
  \bibnamefont{{\em{et~al.}},}
  \bibinfo{journal}{Phys. Rev. Lett.} \textbf{\bibinfo{volume}{79}},
  \bibinfo{pages}{1738} (\bibinfo{year}{1997}).

\bibitem[{\citenamefont{Takanaka}(1975)}]{Takanaka75}
\bibinfo{author}{\bibfnamefont{K.}~\bibnamefont{Takanaka}},
  \bibinfo{journal}{Phys. Stat. Sol. B} \textbf{\bibinfo{volume}{68}},
  \bibinfo{pages}{623} (\bibinfo{year}{1975}).

\bibitem[{\citenamefont{Teichler}(1975)}]{Teichler75}
\bibinfo{author}{\bibfnamefont{H.}~\bibnamefont{Teichler}},
  \bibinfo{journal}{Phys. Stat. Sol. B} \textbf{\bibinfo{volume}{72}},
  \bibinfo{pages}{211} (\bibinfo{year}{1975}).

\bibitem[{\citenamefont{Belashchenko et~al.}(1999)\citenamefont{Belashchenko, {van~Schilfgaarde},
  and Antropov}}]{Belashchenko01}
\bibinfo{author}{\bibfnamefont{K.~D.} \bibnamefont{Belashchenko}}
   \bibnamefont{{\em{et~al.}},}
  \bibinfo{journal}{Phys. Rev. B} \textbf{\bibinfo{volume}{64}},
  \bibinfo{pages}{092503} (\bibinfo{year}{2001}).

\bibitem[{\citenamefont{An and Pickett}(2001)\citenamefont{An and Pickett}}]{An01}
\bibinfo{author}{\bibfnamefont{J.~M.} \bibnamefont{An}}  \bibnamefont{and}
  \bibinfo{author}{\bibfnamefont{W.~E.} \bibnamefont{Pickett}},
  \bibinfo{journal}{Phys. Rev. Lett.} \textbf{\bibinfo{volume}{86}},
  \bibinfo{pages}{4366} (\bibinfo{year}{2001}).

\bibitem[{\citenamefont{Choi et~al.}(2001)\citenamefont{Choi, Roundy,
  Sun, Cohen, and Louie}}]{Choi01}
\bibinfo{author}{\bibfnamefont{H.~J.} \bibnamefont{Choi}}
  \bibnamefont{{\em{et~al.}},}
  (\bibinfo{year}{2001}), cond-mat/0111182.

\end{thebibliography}
\end{document}